\documentclass[12pt]{article}
\input{epsf}
\begin{document}
%
%
\def\mevc {\ifmmode {\rm MeV}/c \else MeV$/c$\fi}
\def\mevcc {\ifmmode {\rm MeV}/c^2 \else MeV$/c^2$\fi}
\def\gevc {\ifmmode {\rm GeV}/c \else GeV$/c$\fi}
\def\gevcc {\ifmmode {\rm GeV}/c^2 \else GeV$/c^2$\fi}
\def\ra   {\rightarrow}
\newcommand{\Bs} {\ifmmode B_{\mbox{\sl s}}^{0}
                       \else $B_{\mbox{\sl s}}^{0}$\fi}
\newcommand{\Ds} {\ifmmode D_{\mbox{\sl s}}^{+}
                       \else $D_{\mbox{\sl s}}^{+}$\fi}
\newcommand{\Dsm} {\ifmmode D_{\mbox{\sl s}}^{-}
                       \else $D_{\mbox{\sl s}}^{-}$\fi}
\newcommand{\dms} {\ifmmode \Delta m_{\mbox{\sl s}} \else 
                           $\Delta m_{\mbox{\sl s}}$\fi}
\newcommand{\eD} {\ifmmode \varepsilon{\cal D}^2 \else 
                          $\varepsilon{\cal D}^2$\fi}
\begin{titlepage}
\vspace{3 ex}
%
%
\begin{center}
{
\LARGE \bf \rule{0mm}{7mm}{\boldmath Future of $B$ Physics at CDF and D\O}\\
}

\vspace{4ex}

{\large
Manfred Paulini~\footnote{Representing the CDF and D\O~Collaboration.} \\
}
\vspace{1 ex}

{\em
Carnegie Mellon University, Pittsburgh, Pennsylvania 15213, U.S.A. \\
}
\vspace{2 ex}
%
%
{\large
Contribution to Panel Discussion on \\
``The Future of Hadron $B$~Experiments''
}
\end{center}

\vspace{2 ex}
%
%
\begin{abstract}
In this contribution to the panel discussion on ``The Future of
Hadron $B$~Experiments'' held at the 8$^{\rm th}$ International Conference
on $B$~Physics at Hadron Machines (Beauty\,2002) at Santiago de Compostela,
Spain, June 17-21, 2002, we explore 
the physics potential for $B$~physics at CDF and D\O\ in five years and
beyond. After a brief introduction to precision flavour physics, we
concentrate our discussion on the future of $CP$~violation by evaluating
the prospects for measuring the CKM angles $\beta$, $\gamma$ and $\alpha$
at the Tevatron Collider experiments CDF and D\O\ by the end of Run\,II.
\end{abstract}

\end{titlepage}

%
\setlength{\oddsidemargin}{0 cm}
\setlength{\evensidemargin}{0 cm}
\setlength{\topmargin}{0.5 cm}
\setlength{\textheight}{22 cm}
\setlength{\textwidth}{16 cm}
\setcounter{totalnumber}{20}
\topmargin -1.0cm

\clearpage\mbox{}\clearpage

\pagestyle{plain}
\setcounter{page}{1}
%
\subsection*{Introduction}
This report contains my contribution to the panel discussion on 
``The Future of Hadron $B$~Experiments'' held at the 8$^{\rm th}$
International Conference 
on $B$~Physics at Hadron Machines (Beauty\,2002) at Santiago de Compostela,
Spain, June 17-21, 2002.
The panel was chaired by Fred Gilman, also current chair
of HEPAP, who charged
the panel members to evaluate the physics potential for
$B$~physics at Hadron Machines in five years and
beyond~\cite{fred_panel}. In my contribution, I discussed the
future of $B$~physics at the two Tevatron Collider experiments CDF and D\O\
at the end of Run\,II.  
More information about $B$~physics prospects at the Tevatron in
Run\,II and beyond can be found in Ref.~\cite{breport}. 

Since the official start of the Fermilab Tevatron Run\,II in March 2001,
much work has gone 
into commissioning the CDF and D\O~detectors. Both experiments were taking
physics 
data and presenting first physics results at the time of this conference in
summer of 2002~\cite{mybeauty}. 
The goal for the first phase of Run\,II (Run\,IIa) is to 
collect a data sample of about 2~fb$^{-1}$ until 2004. After a short
shutdown of about 6 months, Fermilab's current plan foresees a high-luminosity
running period of the Tevatron with an integrated luminosity of possibly
$\sim10$~fb$^{-1}$ delivered until the turn-on of the LHC in 2007. The most
important upgrades for CDF and D\O\ consist of replacing their silicon
vertex detectors by more radiation tolerant devices to allow for
collecting 10-15~fb$^{-1}$ of $p\bar p$~collision data.

\subsection*{Toward Precision Flavour Physics}

\begin{figure}[tbp]
\centerline{
\epsfysize=6.9cm
\epsffile{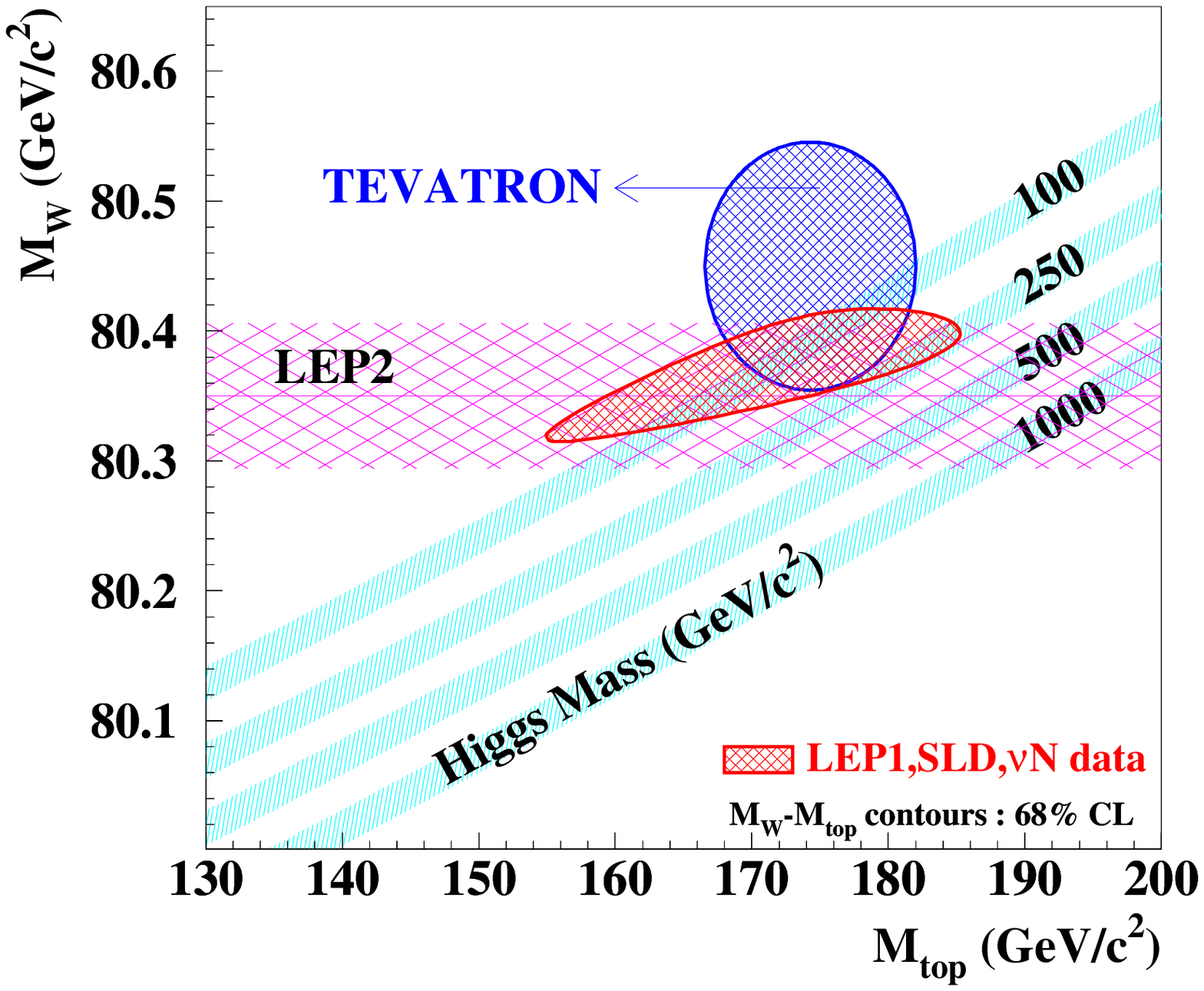}
\epsfysize=6.9cm
\epsffile{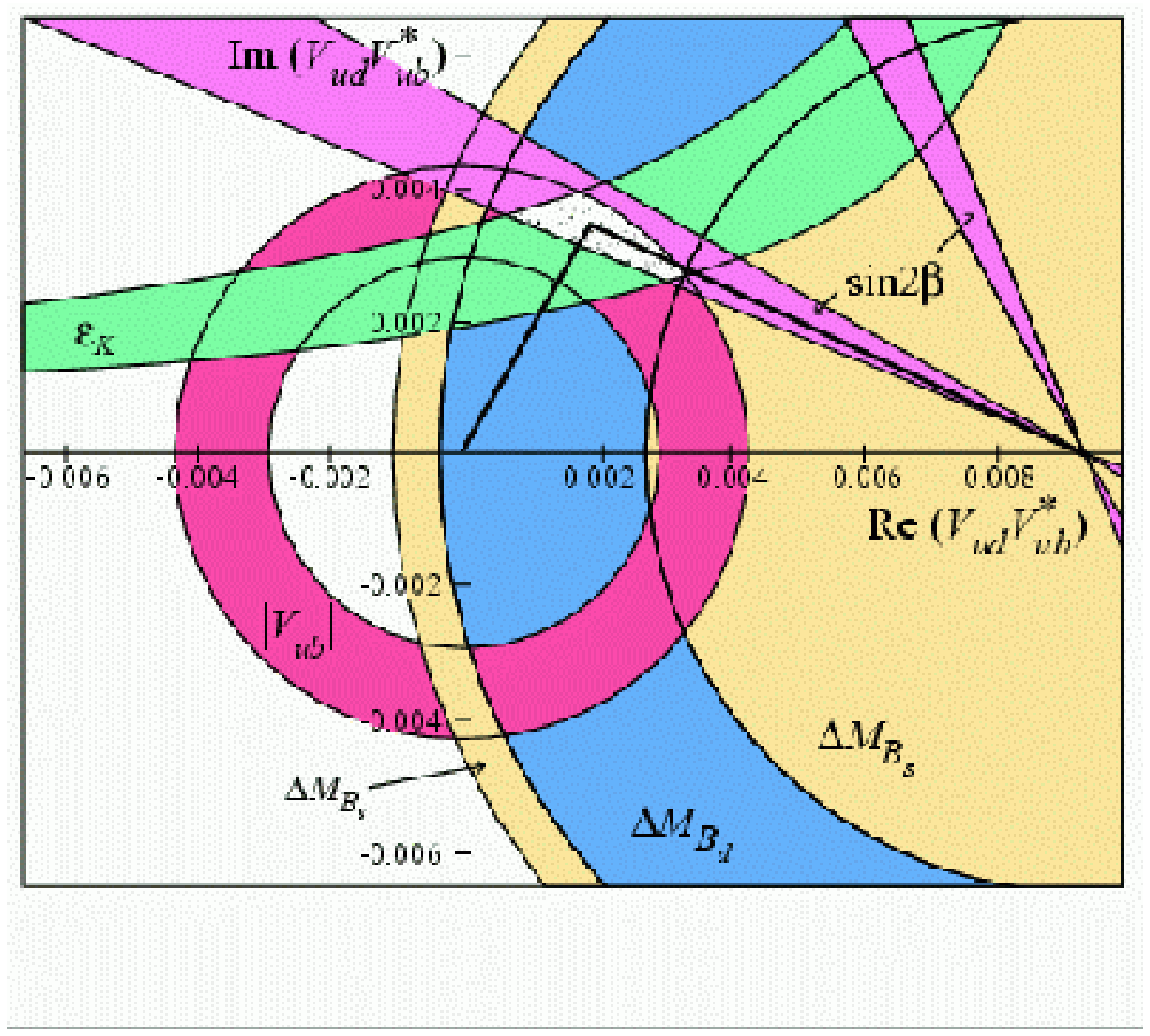}
\put(-410,175){\large\bf (a)}
\put(-210,165){\large\bf (b)}
}
\caption{
(a) Constraints on the Higgs boson mass as of 1999 from measurements of $M_W$
and $M_{\rm top}$ from the Tevatron, LEP\,2 and indirect results. 
(b) Constraints on the position of the apex of the unitarity triangle 
from $|V_{ub}|$, $B$~mixing, $\epsilon$ and $\sin2\beta$ as of 2002
(from Ref.~\cite{ckm_pdg_2002}).
}
\label{mw_ckm}
\end{figure}

In the 1990's, particle physics was dominated by a decade of precision
electroweak physics with measurements from LEP, SLD, the Tevatron and
various fixed target experiments. The progress made in that decade can be
summarized in the constraints on the Higgs boson mass as known in 1999. In
the famous $M_W$ versus $M_{\rm top}$ plane, the Tevatron and LEP
measurements on $M_W$, the CDF and D\O\ results on $M_{\rm top}$ as well as  
constraints from various indirect measurements are displayed in
Figure~\ref{mw_ckm}(a). As indicated, the Standard Model (SM) prefers a light
Higgs Boson mass. Ten years earlier, in 1989, the top quark was not yet
discovered and the $W$~boson mass was only known with a precision of about
1.5~\gevcc. Enormous progress has been made in this decade of precision
electroweak physics.

With the turn-on of the $e^+e^-$~$B$~factories in 1999, the physics
community started to 
talk about particle physics moving from a decade of precision electroweak
physics toward a decade of precision flavour
physics. A lot of progress in understanding flavour physics has already been
achieved in the past 4-5 years. In 1998, we had no information on
$CP$~violation in the $B$~system which was discovered in 2001~\cite{cpb}, 
and the constraints from $|V_{ub}|$, $B$~mixing and $CP$~violation in the
kaon system were quite coarse. Much progress has been made since then as
can be seen in Figure~\ref{mw_ckm}(b) showing the constraints on the
position of 
the apex of the unitarity triangle from $|V_{ub}|$, $B$~mixing, $\epsilon_K$
and $\sin2\beta$. This plot has been taken from the 2002 Review of Particle
Properties~\cite{ckm_pdg_2002}.

\begin{figure}[tbp]
\centerline{
\epsfxsize=8.0cm
\epsffile{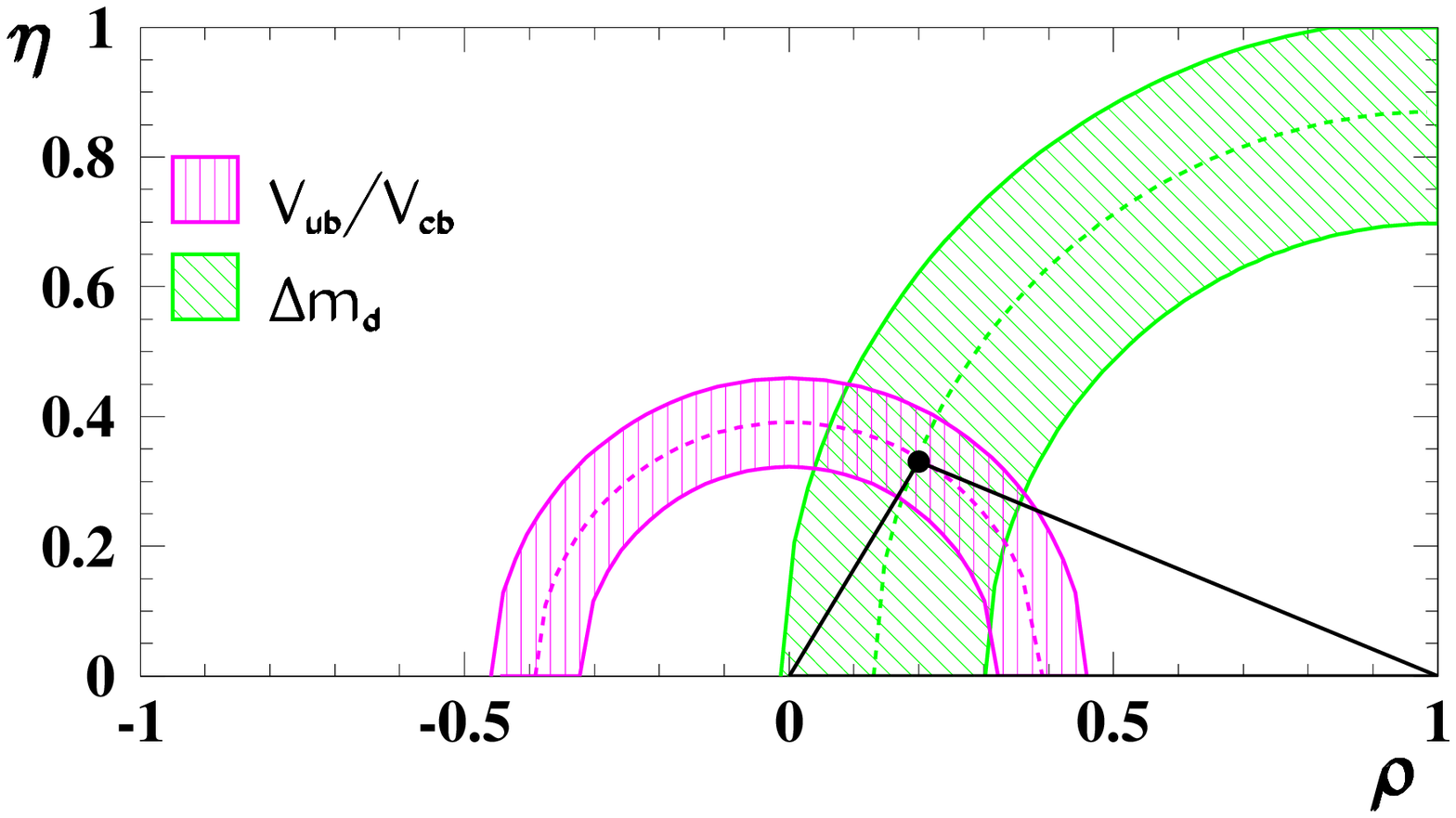}
\epsfxsize=8.0cm
\epsffile{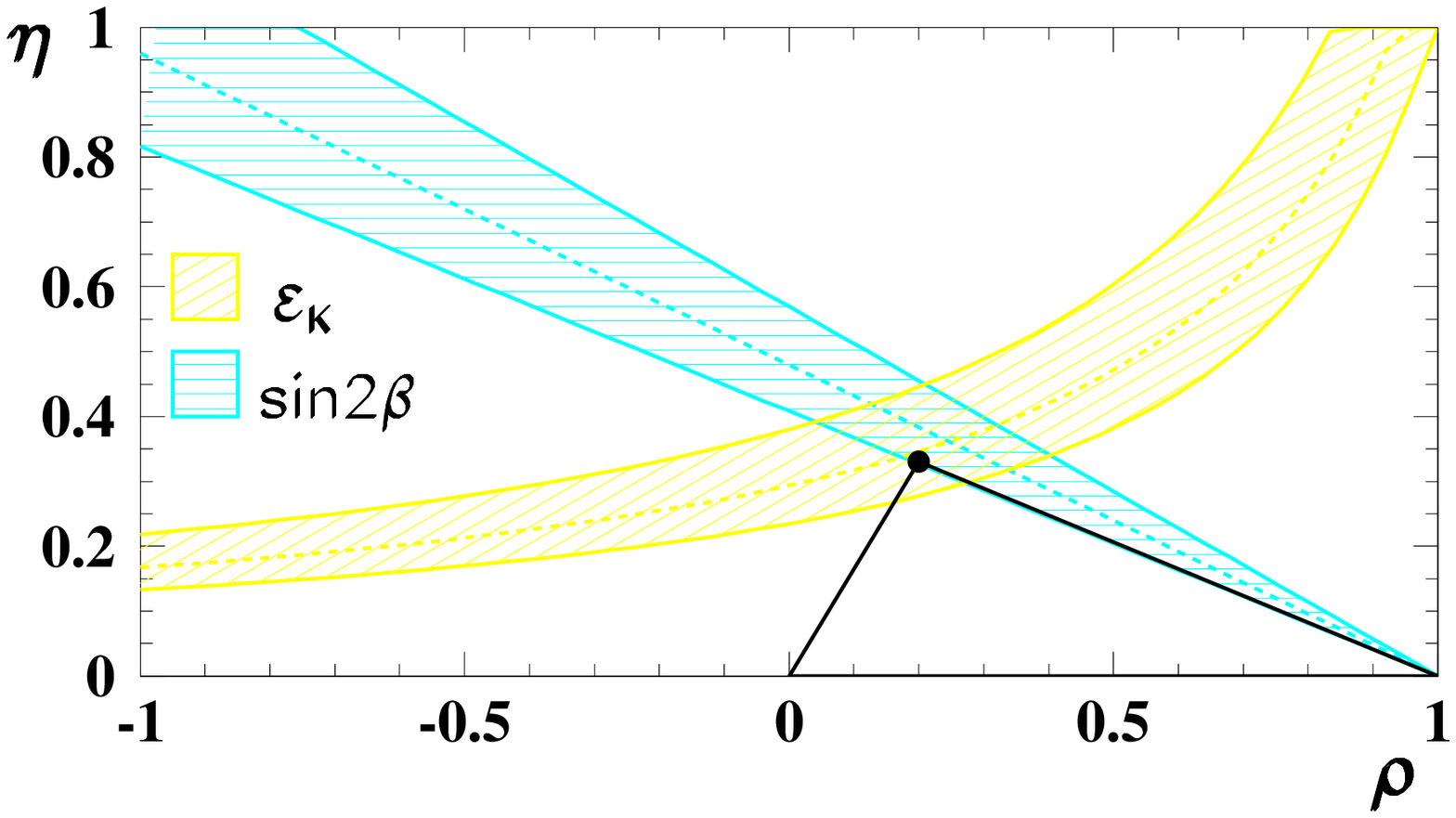}
\put(-375,105){\large\bf (a)}
\put(-120,105){\large\bf (b)}
}
\caption{
Illustration of the two independent tests of the CKM unitarity triangle 
through (a) rate measurements ($|V_{ub}|$ and $B$~mixing) and
(b) $CP$~violation ($\epsilon_K$ and $\sin2\beta$).
}
\label{ckm_rate_cp}
\end{figure}

According to the chair of this panel on ``The Future of Hadron
$B$~Experiments'', Fred Gilman, for the first time there exist now 
two independent tests of the CKM mechanism in the Standard
Model~\cite{fred_panel}. Rate 
measurements of $|V_{ub}|$ as well as $B^0$ and \Bs~mixing
constrain the position of the apex of the unitarity triangle as
illustrated in Figure~\ref{ckm_rate_cp}(a). The second independent test
comes from $CP$~violation in the kaon system ($\epsilon_K$) and the
$B$~system ($\sin2\beta$) as displayed in
Figure~\ref{ckm_rate_cp}(b). Both tests through rate measurements and
$CP$~violation are in striking agreement as illustrated in
Figures~\ref{ckm_rate_cp} and \ref{mw_ckm}(b). 

\subsubsection*{A Look into the Future}

\begin{figure}[tbp]
\centerline{
\epsfxsize=11.0cm
\epsffile{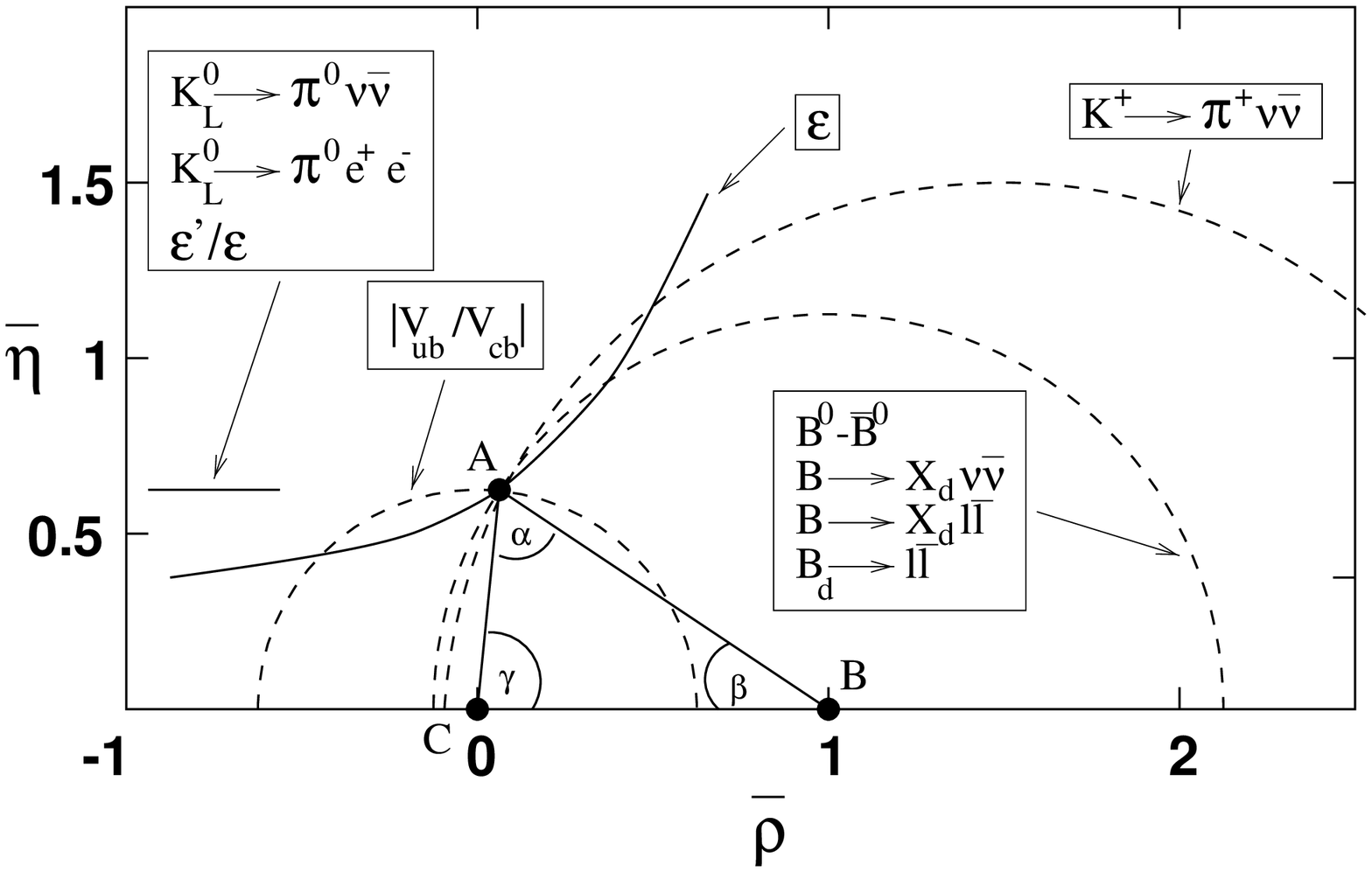}
}
\caption{
An ideal CKM unitarity triangle indicating the different constraints from
$K$~decays and $B$~physics (from Ref.~\cite{buras2001}).
}
\label{ckm_buras}
\end{figure}

When the ``wise (wo)man'' is asked to have a look into the future of
$B$~physics at hadron machines, (s)he might do two things. First, (s)he
might take a step back and recommend to get a ``grand view''. For the
topic of this panel discussion, the ``grand picture'' is presented in
Figure~\ref{ckm_buras} taken from Ref.~\cite{buras2001}. In this diagram 
an ideal view of the CKM unitarity triangle is given indicating the
different constraints from $K$~decays ($\epsilon_K$,
$\epsilon^{\prime}/\epsilon$, 
$K^0_L \ra \pi^0\nu\bar\nu$ and $K^+\ra\pi^+\nu\bar\nu$) as well as 
$B$~physics ($|V_{ub}|$, $B$~mixing, rare decays and $CP$~violation in form
of the angles $\alpha$, $\beta$ and $\gamma$).
The second advice the wise (wo)man might offer is to look at your older
friends. If you want to know how you'll appear in five years from now, look
at your 
older friends and see what they are right now. The older friend of
$B$~physics is kaon physics. Currently, the virtue of kaon experiments is
to test the Standard Model by measuring
modes that are theoretically clean such as   
$K^0_L \ra \pi^0\nu\bar\nu$ and $K^+\ra\pi^+\nu\bar\nu$. Thus, the goal of
$B$~physics in five years and beyond should be to test the SM
by concentrating on theoretically clean modes. These will also allow to
find new physics in an unambiguous way. In the remainder of this paper, we
will therefore focus on testing $CP$~violation through theoretically clean  
$B$~decay modes.

\subsection*{The Future of \boldmath{$CP$}~Violation}

In the following, we assume a data sample of 15~fb$^{-1}$ for the quoted
prospects which would correspond to the luminosity that can possibly be
reached by 
the end of the Tevatron Run\,II. We will mainly concentrate on prospects
determined at CDF as they were more accessible than prospects from D\O.

\subsubsection*{The Angle \boldmath{$\beta$}
from \boldmath{$B^0\ra J/\psi K^0_S$}}

CDF's most important $B$~physics goal in Run\,II is the study of
$CP$~violation in the $B$~system. The golden decay 
$B^0\ra J/\psi K^0_S$ 
is the mode which all experiments will use to obtain precision
measurements of $\sin2\beta$. For Run\,IIa with 2~fb$^{-1}$ of data, CDF
expects $\sim$20,000 fully reconstructed $B^0\ra J/\psi K^0_S$ decays. With
an expected effective tagging efficiency of $\eD\sim9.1\%$, CDF will
measure $\sin2\beta$ with an uncertainty of about 0.05. D\O\ expects a
similar precision. 
The systematic uncertainty on $\sin2\beta$ is dominated by the uncertainty
on the dilution which is determined by large control samples of $J/\psi
K$~data. Thus this uncertainty scales with statistics. Since we do not
see a limiting systematic uncertainty, the precision on $\sin2\beta$
will scale with the integrated luminosity resulting in an uncertainty of
0.02 in 15~fb$^{-1}$ competitive with expected measurements at the
$e^+e^-$~$B$~factories. 

\subsubsection*{\boldmath{$CP$} Asymmetry in \boldmath{$\Bs\ra J/\psi\phi$}}

While the $CP$~asymmetry in $B^0\ra J/\psi K^0_S$ measures the weak phase
of the CKM matrix element $V_{td}$, $CP$~asymmetry in 
$\Bs\ra J/\psi\phi$
measures the weak phase of the CKM matrix element $V_{ts}$.
The latter $CP$~asymmetry is
expected to be very small in the Standard Model, on the order of a few
percent. In the context of testing the Standard Model, this mode has the
same fundamental importance as measuring $\sin2\beta$ but is most
accessible at a hadron collider. An observation of a large $CP$~asymmetry
would be a clear signal of new physics.  

In Run\,I, CDF's yield of $J/\psi\phi$ was about 40\% of $J/\psi K^0_S$
which results in an expectation of about 8000~$J/\psi\phi$ events in
Run\,IIa. The magnitude of the $CP$~asymmetry in $\Bs\ra J/\psi\phi$ is
modulated by the frequency of \Bs~mixing. This requires to resolve
\Bs~oscillations. There is an additional complication in this decay mode,
if the $J/\psi\phi$~final state is not a pure $CP$~eigenstate. In this case,
an angular analysis is necessary to determine the mixture of $CP$~even and
$CP$~odd states in this decay channel. 
With 15~fb$^{-1}$ of data in Run\,II,  
CDF expects a resolution of 0.03-0.06 on the
$CP$~asymmetry in $\Bs\ra J/\psi\phi$ for $\dms\sim20$~ps$^{-1}$ 
depending on the $CP$~content of the final state.

\subsubsection*{\boldmath{$CP$} Asymmetry in 
\boldmath{$\Bs\ra J/\psi\eta^{(\prime)}$}}

\begin{figure}[tbp]
\centerline{
\epsfxsize=7.0cm
\epsffile{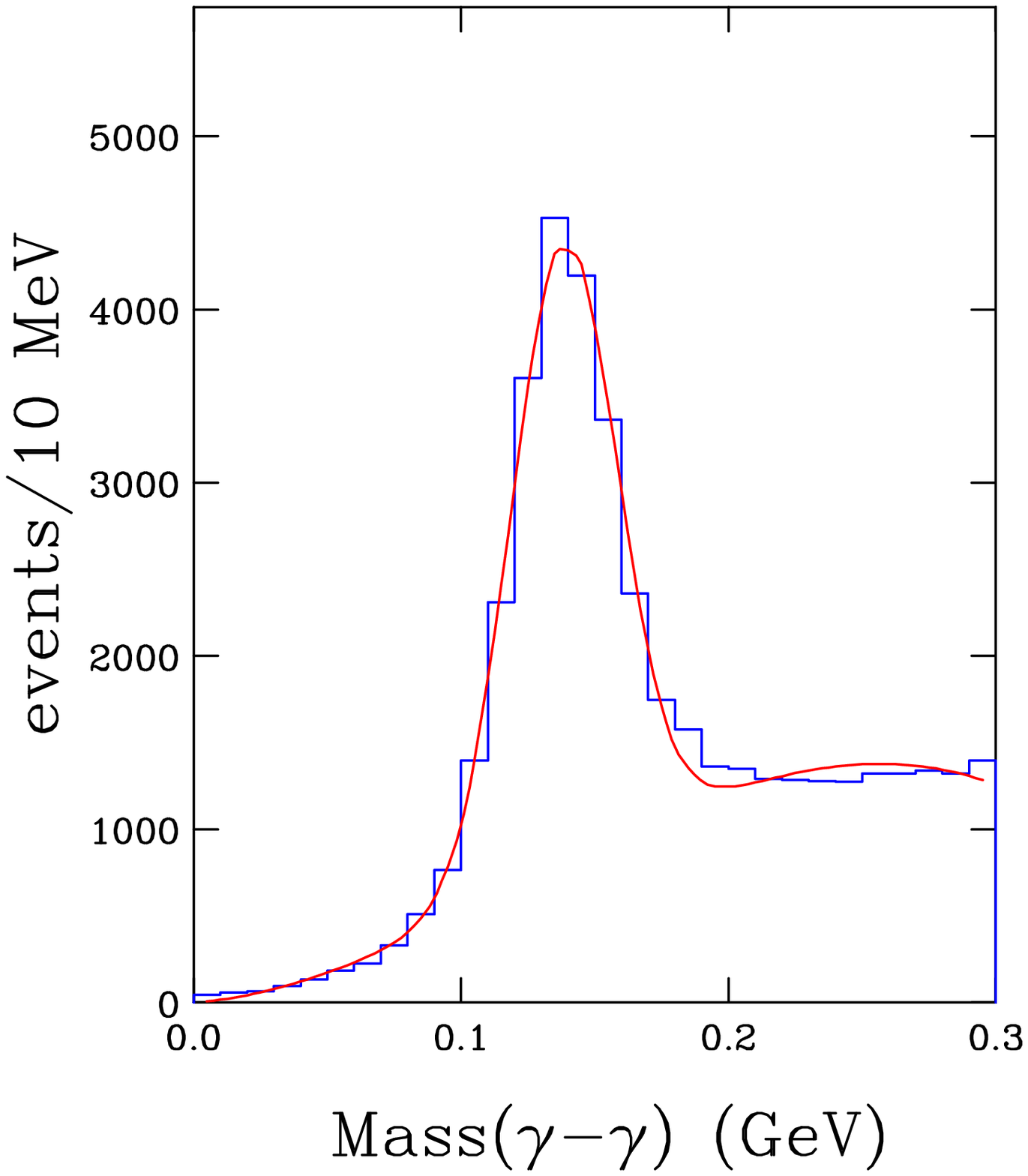}
\epsfxsize=7.0cm
\epsffile{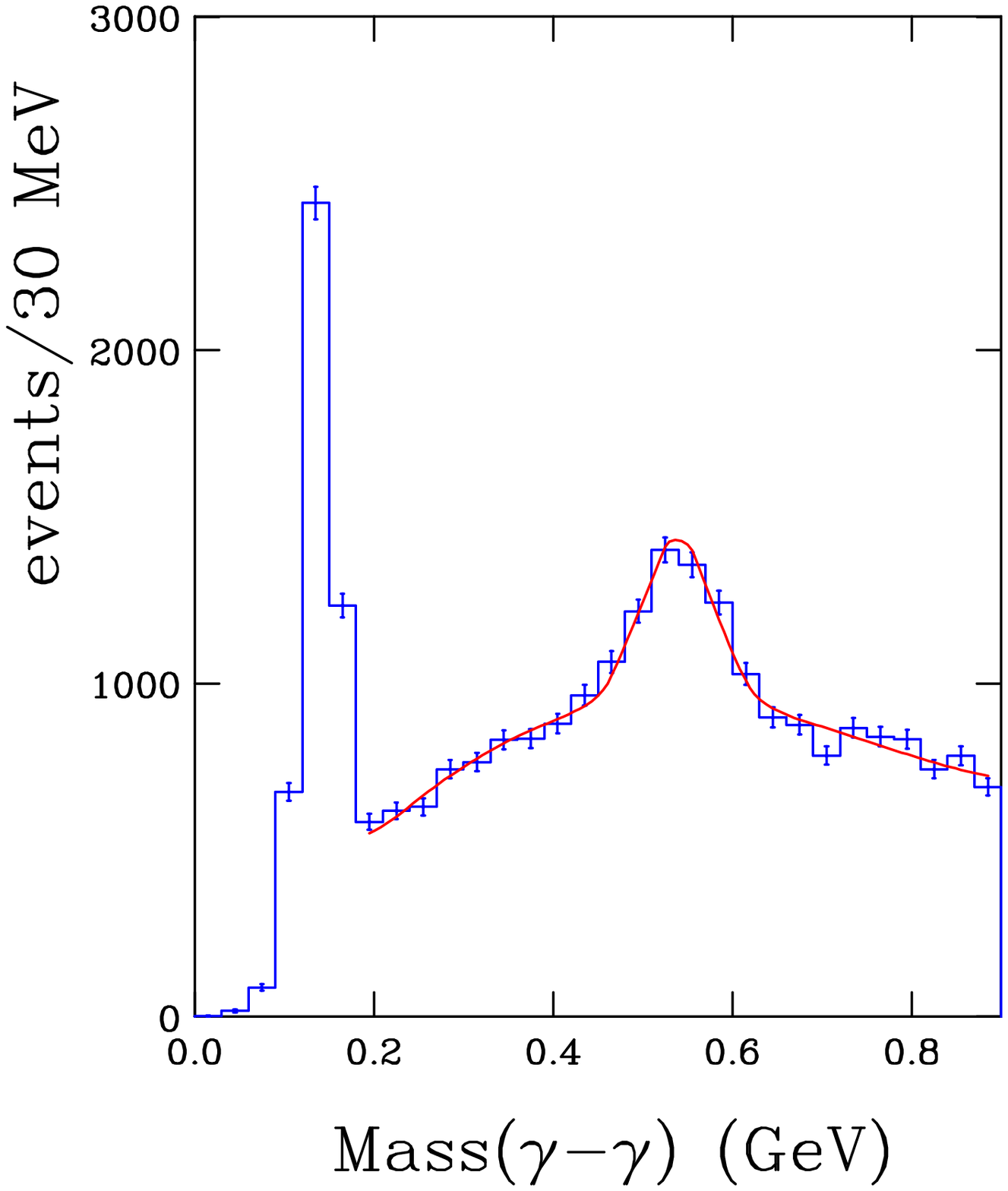}
\put(-350,200){\large\bf (a)}
\put(-50,200){\large\bf (b)}
}
\caption{
Invariant diphoton mass distribution showing 
(a) $\pi^0\ra\gamma\gamma$ and (b) $\eta\ra\gamma\gamma$ 
signals in CDF Run\,I data.
}
\label{cdf_psieta}
\end{figure}

Measuring the $CP$~asymmetry in
$\Bs\ra J/\psi\eta^{(\prime)}$ 
decays is very similar to measuring it in 
$\Bs\ra J/\psi\phi$ with two differences. First, the 
$J/\psi\eta$ and $J/\psi\eta^{\prime}$ final states are
$CP$~eigenstates. Therefore no angular analysis is needed and no
degradation of the $CP$~asymmetry occurs. Second, the presence of  
photons in the final state makes these modes more difficult to detect at
CDF since the CDF calorimeter was not designed to measure low energy
photons with good resolution. However, CDF is capable of detecting these
signals as shown in Figure~\ref{cdf_psieta}. Here, in the
invariant diphoton mass spectrum clear signals of
$\pi^0\ra\gamma\gamma$ and $\eta\ra\gamma\gamma$ are 
observed in CDF Run\,I data.

Scaling from the expected number of $B^+\ra J/\psi K^+$ events, the rate
of $B^0$ to \Bs~production and the expected relative branching ratios, CDF
expects about 8000~$\Bs\ra J/\psi\eta$ events in Run\,II. Studies of
$J/\psi$~events in Run\,I indicate that a mass resolution of 40~\mevcc\
and a signal-to-background ratio of 1:2 appears achievable. For 
$\dms\sim 20$~ps$^{-1}$ and a proper time resolution of
$\sigma_t\sim0.045$~ps, CDF expects to measure the $CP$~asymmetry in 
$\Bs\ra J/\psi\eta$ with an uncertainty on the order of 0.1.

\subsubsection*{The Angle \boldmath{$\gamma$} from 
\boldmath{$\Bs\ra\Dsm K^+$}}

A good candidate to determine the CKM angle $\gamma$ is the decay mode 
$\Bs\ra\Dsm K^+$ measuring $\sin\gamma$. In this mode, $CP$~violation occurs
via interference of the quark level processes $b\ra c\bar us$ and $b\ra
u\bar cs$ through direct and mixed decays. Since \Bs~oscillations are
expected to 
have a small $CP$~violating phase, the relative weak phase of this decay is
${\rm e}^{i\gamma}$. Penguin contributions are expected to be small but
there is a strong phase $\delta$ present which cannot be reliably
calculated with present theoretical techniques and needs to be extracted
from data. 
The time dependent decay rates for all four processes 
$\Bs/\bar \Bs\ra D_{\mbox{\sl s}}^{\mp}K^{\pm}$ are fitted with a two-fold
ambiguity in $\delta$ and $\gamma$. 

The reduction of backgrounds, in
particular physics backgrounds from  the Cabibbo allowed process 
$\Bs \ra \Dsm \pi^+$, is the primary
challenge for CDF in extracting the $\Bs\ra\Dsm K^+$ signal.
Exploiting the $\Dsm K^+$ invariant mass as well as d$E$/d$x$ information of
the final state particles, studies at CDF show that a
signal-to-background ratio of 1/6 can be achieved 
and a signal of
850 $\Bs\ra\Dsm K^+$ events can be expected in 2~fb$^{-1}$. 
Thus, an initial measurement
of $\gamma$ should be possible at CDF in the beginning of Run\,II.
Within the first 2~fb$^{-1}$ of data, the expected error on
$\sin(\gamma\pm\delta)$ is
0.4 to 0.7 depending on the assumed background levels.
By the end of Run\,II, an uncertainty near $0.2$ for $\gamma$ may be
achievable. 

\subsubsection*{The Angle \boldmath{$\gamma$} from 
\boldmath{$B^-\ra D^0 K^-$}}

In a similar manner, the angle $\gamma$ can be determined from the decays 
$B^-\ra D^0 K^-$ and $B^-\ra \bar D^0 K^-$ with $D^0/\bar D^0\ra
K^{\mp}\pi^{\pm}$. The advantage is that these modes are self-tagging and no
time-dependent measurement is necessary. One needs to just measure
branching fractions of these decays. However, the decay 
$B^-\ra \bar D^0 K^-$
is particularly problematic due to the small expected branching ratio. 
All these decay modes have significant physics and combinatoric
backgrounds that must be reduced to acceptable levels to make this method
feasible. 
CDF expects to collect a small sample of about 100 signal 
candidates with the two-track hadronic trigger in 2~fb$^{-1}$.
There is optimism that the physics
background can be brought down to the same level as the signal, but there
could be considerable combinatoric background.
If the combinatoric background can also be reduced to a level
comparable to the signal, CDF would be in the position to measure 
$\gamma$ with an uncertainty in the order of 10-20$^{\circ}$
in 2~fb$^{-1}$. At this point it is not clear whether this can be scaled to
a 5$^{\circ}$ measurement on $\gamma$ with 15~fb$^{-1}$ at the end of
Run\,II. 

\subsubsection*{The Angle \boldmath{$\alpha$}}

To date, there are only two methods considered to be clean extractions
of the $CP$~phase~$\alpha$ from $B$~decays. Each method has its own
particular difficulties. Originally it was suggested to use the time
dependent $CP$~asymmetry in $B^0\ra \pi^+\pi^-$ to obtain $\alpha$. In
order to remove the ``penguin pollution'', it is necessary to perform an
isospin analysis of $B\ra\pi\pi$~decays~\cite{gronau_london} including the
decay mode $B^0\ra\pi^0\pi^0$ which is difficult to measure.
The second method is the Dalitz-plot analysis of $B^0\ra\rho\pi\ra
\pi^+\pi^-\pi^0$ decays~\cite{snyder_quinn}. The problem here is to
understand the continuum background and the correct description of
$\rho\ra\pi\pi$ decays. 

In a recent paper by London and Datta~\cite{londonkk}, a new method to
measure the $CP$~phase~$\alpha$ has been suggested using decays 
$B^0/\Bs\ra K^{(*)}\bar{K}^{(*)}$. 
Because the branching ratios for 
$B\ra K^{(*)}\bar{K}^{(*)}$ are rather small, ${\cal O}(10^{-6})$, and
because \Bs~decays are involved, this method is most appropriate for
hadron colliders, in particular since no $\pi^0$~detection is needed.
The basic idea is to consider the pure $b\ra d$~penguin decays 
$B^0\ra K^0\bar K^0$ and
$B^0\ra K^*\bar K^*$ and relate their
time dependent decay rate to the corresponding \Bs~decay modes into 
$K^0\bar K^0$ and $K^*\bar K^*$ assuming $U$-spin symmetry.
Using a double ratio in which the $SU(3)$-breaking effects largely cancel,
the theoretical uncertainties are estimated to be at most
5\%~\cite{londonkk}.  
The potential weakness of this method is an up to 16-fold ambiguity in
extracting $\alpha$ 
which can be reduced by considering other 
$K^{(*)}\bar K^{(*)}$ final states.
This constitutes a promising method for a potentially clean measurement of
$\alpha$. CDF expects to collect about 100-200
$B\ra K^{(*)}\bar{K}^{(*)}$ decays with its hadronic track trigger in
1~fb$^{-1}$. 

\subsection*{Conclusion}

In this contribution to the panel discussion on 
``The Future of Hadron $B$~Experiments'' held at the Beauty\,2002
conference, we made an attempt to discuss the question ``Quo vadis,
$B$~physics?'', where is $B$~physics going in five years and beyond by 
representing the $B$~physics prospects at CDF and D\O.
There exists a long laundry list of modes to measure at the $B$~factories,
the Tevatron and $3^{\rm rd}$~generation $B$~experiments (BTeV, LHCb). In
the kaon system two clean tests of the Standard Model have been identified, 
$K^0_L \ra \pi^0\nu\bar\nu$ and $K^+\ra\pi^+\nu\bar\nu$, but it is not
clear which are the ``smoking gun modes'' in the
$B$~system. Precision measurements of $\sin2\beta$, $\Bs\ra J/\psi\phi$ and
possibly $B\ra K^{(*)}\bar{K}^{(*)}$ decays are likely good candidates.

Going back to looking at the ``grand picture'', the relation of the CKM
matrix to the quark mass hierarchy, flavour physics, the Higgs mechanism
and electroweak 
symmetry breaking, it would be desirable to have a well defined path with
clean Standard Model tests to be performed in the $B$~system. Since this
path is not obvious at this point, the only answer can be to continue
strengthening the
planned experimental efforts (Run\,IIb, BTeV, LHCb) to ask questions to
nature by doing experiments in order to test the flavour
sector of the Standard Model until it breaks.
Let me conclude with a quote from Woody Allen which does not only relate to
the lack of access to divine counseling but also to the funding situation 
in US
high energy physics: {\it ``If only God would give me some clear sign! Like
making a large deposit in my name at a Swiss bank.''}
    



\begin{thebibliography}{99}

\bibitem{fred_panel} F.~Gilman, these proceedings and private communication.

\bibitem{breport} K.~Anikeev {\it et al.},
FERMILAB-PUB-01-197 (2001) 
[arXiv:hep-ph/0201071].

\bibitem{mybeauty} M.~Paulini (for the CDF Collaboration), these
		proceedings; \\
		FERMILAB-CONF-03/010-E (2002).

\bibitem{cpb} 
B.~Aubert {\it et al.}  [BABAR Collaboration],
Phys.\ Rev.\ Lett.\  {\bf 87}, 091801 (2001); \\
K.~Abe {\it et al.}  [Belle Collaboration],
Phys.\ Rev.\ Lett.\  {\bf 87}, 091802 (2001). 

\bibitem{ckm_pdg_2002} F.~Gilman, K.~Kleinknecht and B.~Renk 
                   ``The Cabibbo-Kobayashi-Maskawa Quark-Mixing Matrix'', in
                   K.~Hagiwara {\it et al.} (Particle Data Group), \\ 
		   Phys.\ Rev.\ D {\bf 66}, 010001 (2002).

\bibitem{buras2001} A.~J.~Buras, [arXiv:hep-ph/0101336].

\bibitem{gronau_london} 
M.~Gronau and D.~London,
Phys.\ Rev.\ Lett.\  {\bf 65}, 3381 (1990).

\bibitem{snyder_quinn} 
A.~E.~Snyder and H.~R.~Quinn,
Phys.\ Rev.\ D {\bf 48}, 2139 (1993); \\
H.~R.~Quinn and J.~P.~Silva,
Phys.\ Rev.\ D {\bf 62}, 054002 (2000)
[arXiv:hep-ph/0001290].

\bibitem{londonkk} 
A.~Datta and D.~London,
Phys.\ Lett.\ B {\bf 533}, 65 (2002)
[arXiv:hep-ph/0105073].

\end{thebibliography}
\end{document}